# Giant magnetoresistance and spin Seebeck coefficient in zigzag α-graphyne nanoribbons

Ming-Xing Zhai[1], Xue-Feng Wang[1,*], and P. Vasilopoulos[2], Yu-Shen Liu[3#], Yao-Jun Dong[1], Liping Zhou[1], Yong-Jing Jiang[4], Wen-Long You[1]

[1]College of Physics, Optoelectronics and Energy, Soochow University, 1 Shizi Street, Suzhou, Jiangsu 215006, China

[2]Concordia University, Department of Physics, 7141 Sherbrooke Ouest, Montréal, QC, Canada, H4B 1R6

[3] College of Physics and Engineering, Changshu Institute of Technology and Jiangsu Laboratory of Advanced Functional materials, Changshu 215500, China

[4]Department of Physics, Zhejiang Normal University, Jinhua, China 231004

* E-mail: xf_wang1969@yahoo.com

[#] E-mail: ysliu@cslg.edu.cn

We investigate the spin-dependent electric and thermoelectric properties of ferromagnetic zigzag α-graphyne nanoribbons (ZαGNRs) using the density-functional theory combined with the non-equilibrium Green's function method. A giant magnetoresistance is obtained in the pristine even-width ZαGNRs and can be as high as $10^6$ %. However, for the doped systems, a large magnetoresistance behavior may appear in the odd-width ZαGNRs rather than the even-width ones. This suggests that the magnetoresistance can be manipulated in a wide range by the dopants on edges of ZαGNRs. Another interesting phenomenon is that in the B- and N-doped even-width ZαGNRs the spin Seebeck coefficient is always larger than the charge Seebeck coefficient, and a pure-spin-current thermospin device can be achieved at specific temperatures.

## 1. Introduction

Graphene's unique electron transport properties make it a promising material for nano-devices[1-4], including the extraordinary high carrier mobility[1,5], anomalous integer Quantum Hall Effect[6,7]. It can be viewed as a zero gap semiconductor with a linear dispersion relation near the Dirac points[1,8]. The charge carriers are described by a two-dimensional (2D) Dirac Hamiltonian without the spin degree of freedom [1] and behave like massless, relativistic fermions with a "speed of light" equal to the Fermi velocity. In addition, its hexagonal symmetric structure contains two equivalent sublattices and results in two equivalent Dirac points, K and K', in the first Brillouin zone. From the perspective of application, we need to open an energy gap. A feasible method is to pattern graphene to nanoribbons[9].

Top-down[10] and bottom-up[11,12] techniques have been developed to fabricate graphene nanoribbons (GNRs) with different widths. Interesting device-orientated properties have been predicted or found in GNRs such as negative differential resistance[10,13-15], current rectification[12,16], giant magnetoresistance[17,18], spin filtering[19], and thermoelectric properties[20]. Specifically, the edge magnetism exists in zigzag GNRs (ZGNRs) though graphene is a nonmagnetic materials[21] and their transversal geometry symmetry can play a key role in determining their transport properties[22,23]. Employing doping and other modification techniques, we can easily manipulate the magnetic properties in ZGNRs and might make them important materials for spintronics.

In the last few years, 2D carbon allotropes have attracted tremendous attention since 2D-graphdiyne, which consists of hexagons connected by linear carbon chains and shows similar energy bands as graphene, has been successfully synthetized by G. Li *et al.*[24]. 2D carbon allotropes and their stable structures were first predicted in 1987[25]. In the next year, Narita *et al.*[26] optimized graphyne and graphdiyne structures and found that the linear carbon (C) chain is the acetylenic linkage (—C≡C—) rather than the ethylene-like one (=C=C=)[26]. Very recently, Enyashin *et al.*[27] published the electronic properties of twelve different 2D carbon allotropes. Malko *et al.*[28] presented a systematic comparison of the electronic structures of graphene, α-graphyne, shown in Fig. 1(a), β-graphyne, and γ-graphyne, and found the same

Dirac cones as in graphene. Moreover, they noticed that γ-graphyne has two self-doped nonequivalent distorted Dirac cones because of its non-hexagonal symmetry[29]. Lately, Ouyang *et al.*[30] investigated the thermal transport in γ-graphyne ribbons and Yue *et al.*[31, 32] reported properties of the electronic structure and transport in zigzag α-graphyne-based nanoribbons (ZαGNRs).

Electron transport is the physics origin of many properties concerning the movement of electron charge and spin in materials under external electric and magnetic fields or around an environment with thermal gradient. Those properties have been widely used in device designing. For example, the electric properties like giant magnetoresistance is the physical mechanism behind the key technologies for many high-density storage devices[33]. And the thermoelectric properties such as the Seebeck effect can be used to convert thermal energy to electric energy or vice versa and produce spin current from thermal gradient[34]. Both the external-bias-induced electric properties and the temperature-difference-induced thermoelectric properties are determined by the low-energy electron excitations and are related to the conductance near the Fermi energy. Nevertheless, they reflect the excitations from different perspectives and have versatile applications. In the classical level, Mott relation reveals an inverse trend between the electric conductivity and the Seebeck coefficient. In the quantum level, however, this classical result may not hold[35]. It is necessary to study both properties of nano materials in a parallel manner where quantum effects dominate the electron transport.

Recently, a giant magnetoresistance is predicted in monohydrogenated zigzag silicone nanoribbons[36,37] of even-width, similar to the case in ZGNRs[17,18]. This is due to the sharp change of conductance near the Fermi energy in different magnetization configurations. Since the Seebeck coefficient is related not only to the magnitude but also to the slope of conductance spectra at the Fermi energy, it may reveal extra information. For example, a thermoelectric measurement has been used to explore the effect of chemical structure on the electronic structure and charge transport[38]. Furthermore, using a latest developed spin-detection technique, Uchida et al. have successfully measured a spin voltage in a metallic magnet subjected to a temperature gradient[34]. This phenomenon, now called the spin Seebeck effect, provides an alternative method to achieving a pure spin current by using the temperature gradient in the absence of the electric fields. This pioneering experiment has inspired many theoretical and experimental studies on the spin thermoelectric effect in various systems[39-45]. Very recently, a perfect spin-filtering effect and large Seebeck effect was achieved in ZGNRs by a non-magnetic edge doping[20].

In this paper, we report a systematical study of the electric and thermoelectric properties in both pristine and edge-doped Z$\alpha$GNRs in the ferromagnetic (FM) state. The geometric structure of a 2D $\alpha$-graphyne is shown in Fig. 1(a) and a two-probe Z$\alpha$GNR of width $n$ (the number of zigzag chains therein), denoted as $n$-Z$\alpha$GNR, is illustrated in Fig. 1(b). The black solid spheres are the C atoms and the small gray ones the hydrogen atoms that passivate their edge $\sigma$ bonds. We show that an impurity atom (in orange) that replaces an edge C atom can make $n$-Z$\alpha$GNRs, with $n$ odd, materials of large magnetoresistance. In contrast, large magnetoresistance appears only in pristine $n$-Z$\alpha$GNRs of even $n$. Furthermore, we predict that pure spin Seebeck effect (with zero charge Seebeck coefficient) can be observed in $n$-Z$\alpha$GNRs of even $n$. In section 2 we present the transport model and some details of the calculations and in section 3 the results. A summary follows in section 4.

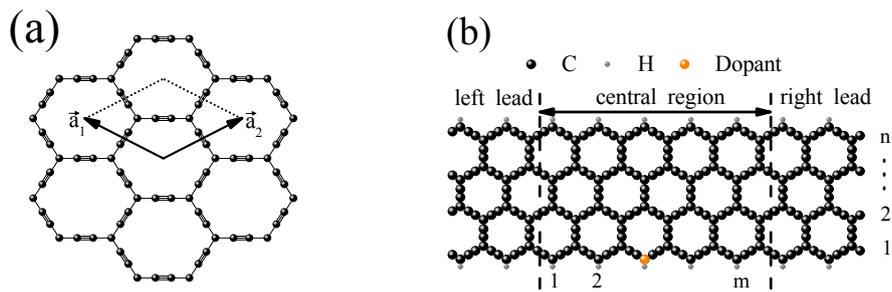

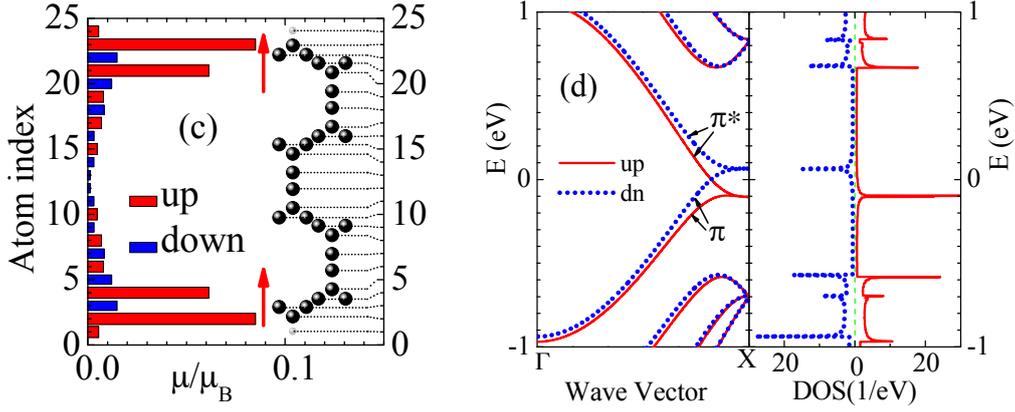

Fig. 1. (Color online) (a) Geometric structure of 2D α-graphyne; $\vec{a}_1$ and $\vec{a}_2$ are basis vectors. (b) A two-probe ZαGNR system of width $n$, denoted as $n$-ZαGNR, is composed of the left and right leads and the central device region of size $n \times m$. Here the width $n$ is defined as the number of parallel zigzag C atom chains in the ribbon and the length $m$ is defined by the number of unit cells in the central region. An impurity atom, in orange, replaces a C (in black) atom on the lower edge. The gray spheres are H atoms. (c) The atomic magnetic moments versus the atom index for atoms in a unit cell of a 4-ZαGNR as shown in the inset. The atom indices (from 1 to 24) refer to the atom rows (from bottom to top in the inset) of the 4-ZαGNR as guided by the dotted lines to the right. The arrows indicate the direction of the magnetization on the edges. The red (blue) bars are for spins up (down). (d) Electronic band structure and density of states (DOS) of spin up (solid curve) and spin down (dotted curve) electrons in the FM state with the Fermi energy at $E=0$.

## 2. Model and method

As shown in Fig. 1(a), the structure of α-graphyne is formed by inserting an acetylenic linkage (—C≡C—) into each C—C bond of graphene [25] and there exist $sp^2$ and $sp$ orbital hybridizations. Similar to GNRs, the ZαGNRs and the armchair α-graphyne nanoribbons also have edges of zigzag and armchair forms, respectively.

To study electronic transport of an $n$-ZαGNR with the dangling edge σ bonds saturated by H atoms, as shown in Fig. 1 (b), we establish a two-probe system by partitioning the ZαGNR into three parts, the central region where an impurity atom may exist and the semi-infinite left and right electrodes. The size of the central region is $n\times m$ with length $m$ the number of the unit cells along the longitudinal direction. There are two magnetization configurations for ferromagnetic electrodes: the magnetization directions of the two electrodes can be parallel (P) or anti-parallel (AP) to each other. Using the Atomistix ToolKits (ATK) package, we have at first optimized the structure geometry using the Newton method with a force tolerance of 0.05 eV/Å and a trust radius of 0.5 Å. The following parameters are obtained: the lengths of the triple and single C−C bonds are 1.23Å and 1.40 Å, respectively; the C−H bond length is 1.10 Å, and the lattice constant 6.98 Å in agreement with Ref. [46]. We choose $m$=5 corresponding to a central region of length 34.9Å, which is long enough to ensure that the left and right electrodes do not couple with each other and also are not affected by the impurity atom.

Our calculations for the electronic structure and transport properties are carried out using the NanoAcademic Device Calculator (NanoDcal) package [47] based on the density functional theory (DFT) combined with the non-equilibrium Green's function (NEGF) method. We employ an exchange-correlation functional in the local density approximation with the Perdew-Zunger (PZ) parameters [48], a real space cutoff energy of 400 Ry and a double-ζ polarization linear-combination-of-atomic orbital basis set for all atoms, and a $k$-point sampling of $1\times1\times100$ grid in the 1D Brillouin zone. A vacuum layer 15Å wide is inserted between the edges and planes of the ZαGNRs in the supercell.

The simulation procedure in NanoDcal is described briefly as follows: The electronic structure of the two electrodes is first calculated to get a self-consistent potential. This potential provides natural real space boundary conditions for the Kohn-Sham effective potential in the central scattering region. Then from the Green's function of this region, we obtain the density matrix and thereby the electronic density. Once the latter is known, the DFT Hamiltonian matrix, which is used to evaluate the Green's function, is computed using the above boundary conditions. This procedure is iterated until the self-consistency is achieved. In the linear response regime, the total conductance of this system is obtained from the Landauer formula [49,50]

$$G_T(\varepsilon) = \sum_\sigma G_\sigma(\varepsilon) = \sum_\sigma (e^2/h)\tau_\sigma(\varepsilon) \tag{1}$$

with $G_\sigma(\varepsilon)$ the conductance of spin $\sigma$ (up↑ or down↓) and the corresponding spin polarization is defined by

$$\eta = (G_\uparrow - G_\downarrow)/(G_\uparrow + G_\downarrow). \tag{2}$$

Here $G_0 = e^2/h$ is the conductance quantum and $\tau_\sigma(\varepsilon)$ the electronic transmission of spin $\sigma$ given by

$$\tau_\sigma(\varepsilon) = Tr[\Gamma_L(\varepsilon)G^r(\varepsilon)\Gamma_R(\varepsilon)G^a(\varepsilon)]_\sigma. \tag{3}$$

The retarded (advanced) Green's function of the central region, $G^{r(a)}(\varepsilon)$, is calculated from the Hamiltonian of this region and the self-energies of the electrodes. The broadening function, $\Gamma_{L(R)}(\varepsilon)$, is evaluated by doubling the imaginary part of the self-energies of the left (right) electrode. The self-energies are computed recursively from the Hamiltonian of infinite electrodes, obtained from an initial bulk calculation of the electrodes [47,51,52].

The tunneling magnetoresistance (MR) for the two-probe system with magnetic electrodes can be evaluated as [33]

$$\text{MR} = \frac{G_T^P - G_T^{AP}}{\text{Min}\{G_T^P, G_T^{AP}\}}, \tag{4}$$

where the linear conductance $G_T^P = G_\uparrow^P + G_\downarrow^P$ and $G_T^{AP} = G_\uparrow^{AP} + G_\downarrow^{AP}$ are the total ones at the Fermi energy in the P and AP magnetization configurations of the electrodes, respectively.

The spin-dependent electric current through the two-probe system is calculated by

$$I_\sigma = \frac{e}{h}\int d\varepsilon[f_\sigma^L(\varepsilon, E_{F\sigma}^L, T^L) - f_\sigma^R(\varepsilon, E_{F\sigma}^R, T^R)]\tau_\sigma(\varepsilon), \tag{5}$$

where $f_\sigma^L$ and $f_\sigma^R$ denote the Fermi-Dirac distribution function $f(\varepsilon, E_F, T) = 1/\{1+\exp[(\varepsilon - E_F)/k_B T]\}$ for spin $\sigma$ in the left (*L*) and right (*R*) electrode, respectively. *T* is the electron temperature and $E_F$ the Fermi energy. From Eq. (5) we see that the spin current $I_\sigma$ can be manipulated by the temperature difference $\Delta T$ or a spin-dependent voltage bias $\Delta V_\sigma$ between the two electrodes. The spin-dependent Seebeck coefficient $S_\sigma$ describes $\Delta V_\sigma$ generated by $\Delta T$ in an open circuit ($I_\sigma = 0$). In the linear response regime, $S_\sigma$ is given by [53-55]

$$S_\sigma = -\lim_{\Delta T \to 0} \frac{\Delta V_\sigma}{\Delta T} = -\frac{1}{eT}\frac{K_{1\sigma}(E_F,T)}{K_{0\sigma}(E_F,T)}, \tag{6}$$

where $K_{\nu\sigma}(E_F,T) = -\int d\varepsilon [\partial f(\varepsilon, E_F, T)/\partial \varepsilon](\varepsilon - E_F)^\nu \tau_\sigma(\varepsilon)$ with $\nu = 0$, or 1. At low temperature, $S_\sigma$ reads approximately[20]

$$S_\sigma \approx -\frac{\pi^2 k_B^2 T}{3e}\frac{\tau'_\sigma(E_F)}{\tau_\sigma(E_F)}. \tag{7}$$

This expression shows that $S_\sigma$ is related not only to the value of $\tau_\sigma(\varepsilon)$ but also to its slope at the Fermi energy. The usual charge Seebeck coefficient is $S_C = (S_\uparrow + S_\downarrow)/2$ whereas the spin Seebeck coefficient is defined as $S_S = (S_\uparrow - S_\downarrow)/2$. The coefficient $S_S$ reflects the ability of the device to produce a spin current induced by a temperature gradient.

### 3. Results and discussions

*A. Spin-dependent energy structure*

There are two spin states for ZαGNRs: the FM state and the antiferromagnetic (AFM) one. In the FM state, both edges are spin-up polarized as illustrated in the inset of Fig. 1(c) for the primitive cell of a 4-ZαGNR while in the AFM state the upper edge is spin-down polarized. The AFM state is the ground state with a lower total energy of 0.01eV per primitive cell in the absence of an external field. The FM state can be easily obtained with the help of an external magnetic field. The edge magnetization originates mainly from the spin splitting of the edge states near the Fermi energy. In Fig. 1(c), we present the magnetic moment of each atom marked by the row number of the 4-ZαGNR in the FM state. The H atoms are polarized in the same direction as their edge and the magnetic moment of the C atoms changes sign alternatively and decreases oscillatorily with the distance from the edge. The atoms with even indexes are much less magnetized. The corresponding electronic band structure and the density of states (DOS) are plotted in Fig. 1(d) for spin up (solid curve) and spin down (dashed curve) electrons with the Fermi energy $E_F$=0. The band structure is spin dependent and shows a metallic behavior. The π and π* bands of the edge states near wave vector point X have almost the same energy and may twist with each other for each spin. The energies of edge states for majority or up (minority or down) spin are 0.1 eV below (0.06 eV above) the Fermi energy.

## B. Conductance in the parallel configuration

For a pristine FM $n$-ZαGNR with the two electrodes magnetized in the same direction, i.e., in the parallel (P) configurations, the conductance spectra can be directly derived from its band structure. In Fig. 2(a) we plot the spin dependent conductance $G_\sigma^P$ for $n$=3, 4, 5, and 6 from top to down along one column. A wide conductance platform of value $G_0$ appears with a possible sharp peak above (below) the Fermi energy for spin-up (down). This corresponds to at least one transport channel for each spin and more channels in the energy range of the band twist as shown in Fig. 1(d) for $n$=4.

When one edge atom C is replaced by a doping atom, conductance dips appear due to the Fano effect arising from the formation of impurity bound states. The positions of the dips in energy indicate the doping type determined by the electrons transferred from the impurity to host atoms. Sharp dips may also appear beside the peaks due to the twist of π and π* bands. In addition, the conductance becomes strongly spin dependent near the Fermi energy due to the breakdown of geometry symmetry in the system though it is only slightly spin polarized at the Fermi energy for all the cases doped by elements of groups III and V (B and N). The spin-up (solid red line) and spin down (dotted blue line) conductance of edge-doped 3-, 4-, 5- and 6-ZαGNRs in the P configuration is presented in Figs. 2 (b)-(c) for doping elements of groups III and V. As can be seen, a dip appears near 0.3 (-0.3) eV for B (N) corresponding to the n type donor (p type acceptor) doping and the dip width narrows as $n$ increases. The electron transfer analysis indicates that the B (N) atom donates (accepts) electrons to (from) the host C atoms. This transition phenomenon of doping type on edge is similar to that in zigzag graphene and silicene nanoribbons [36,56].

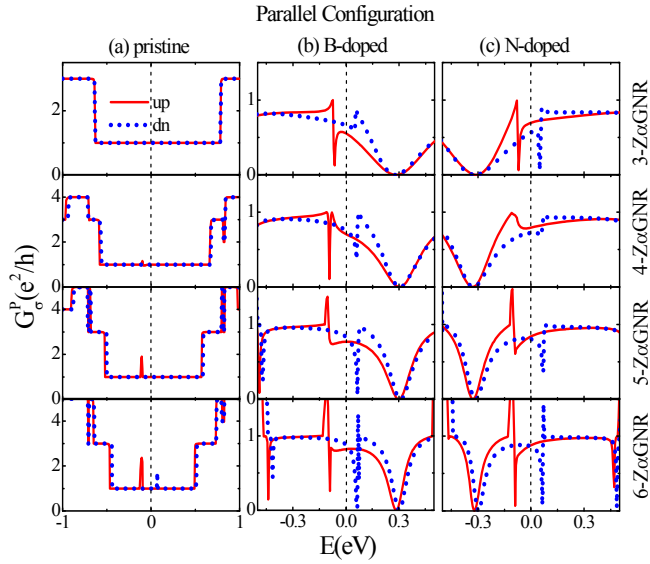

Fig. 2. (Color online) Spin dependent conductance $G_\sigma^P$ in the P configuration. The panels in the 1-4 rows are for 3-, 4-, 5-, 6-ZαGNRs, respectively and those in the three columns from left to right are for pristine, B-, and N-doped ZαGNRs, respectively. The solid red (dotted blue) line represents the spin up (down).

### C. Conductance in the anti-parallel configuration

In the anti-parallel (AP) magnetization configuration of the two electrodes, the conductance $G_\sigma^{AP}$ shows a strong dependence on the parity of ZαGNRs. In Figs. 3(a)-(c) we present the spin-dependent conductance spectra of $n$-ZαGNRs with different width $n$. For odd width pristine $n$-ZαGNRs, the conductance spectrum has a plateau without gap at $E_F$. In contrast, for even width pristine $n$-ZαGNR, a gap of 0.15 eV emerges at $E_F$. This result suggests that odd width pristine ZαGNRs behave like metal while even width pristine ZαGNRs show semiconductor characteristics in the AP configuration. As a result, giant magnetoresistance appear in even-width ZαGNRs similar to that observed in pristine graphene [17] and silicene [37] nanoribbons.

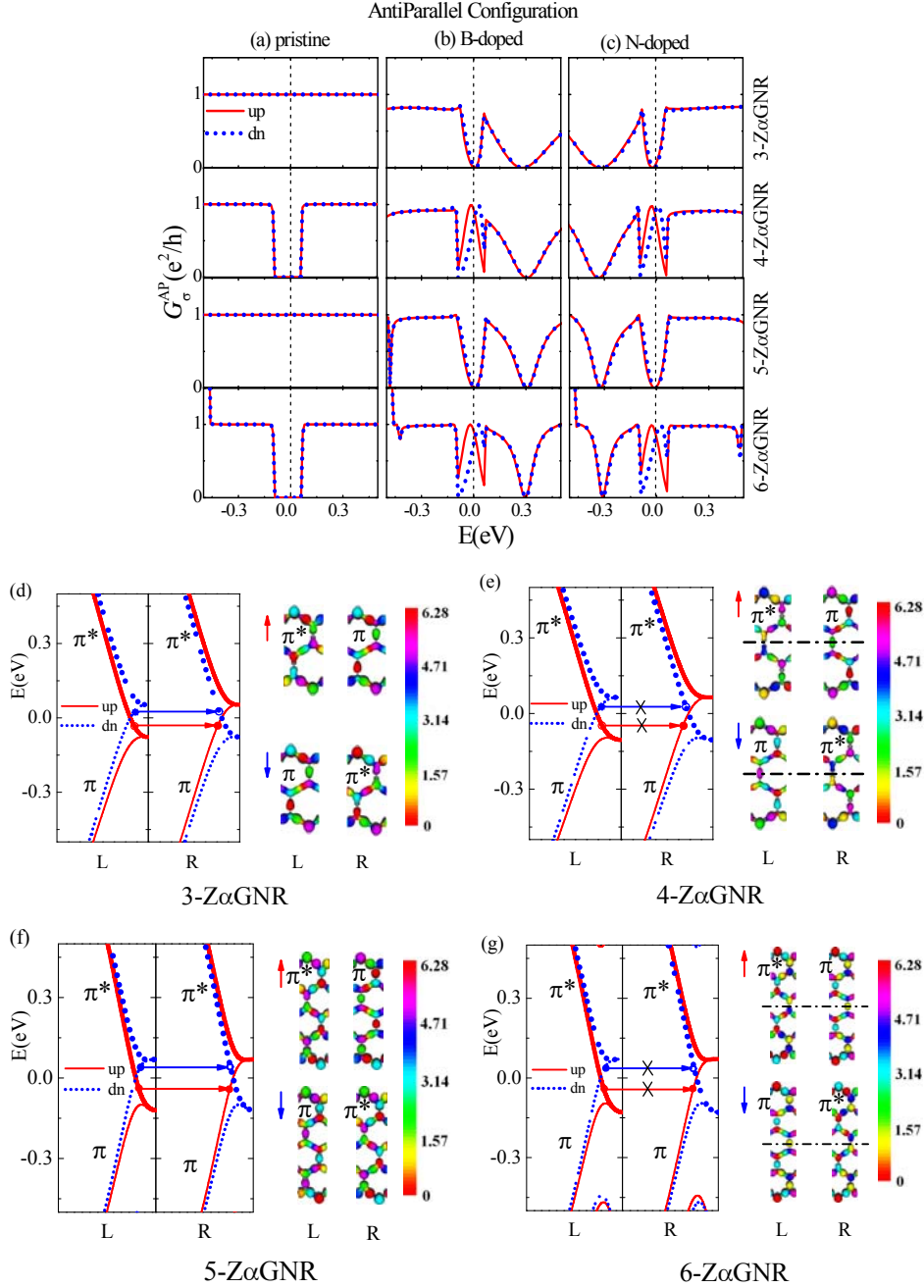

Fig. 3. (Color online) (a)-(c) Spin-dependent conductance $G_\sigma^{AP}$ in the AP configuration. The panels in the 1-4 rows are for 3-, 4-, 5-, 6-ZαGNRs, respectively and those in the columns (a)-(c) from left to right are for pristine, B-, and N-doped ZαGNRs, respectively. The curves are marked

as in Fig. 2. (d)-(g) Band structures of the left (L) and the right (R) electrode for FM 3-, 4-, 5- and 6-ZαGNR, respectively, in the AP configuration. The arrows (crossed arrows) indicate the allowed (forbidden) tunneling processes near the Fermi energy. The corresponding wave functions, with phases marked by colors, of π and π* subbands at the Fermi energy are plotted beside. The horizontal black dash-dotted lines denote the mirror plane.

We note that $n$-ZαGNRs have similar band structures for even and odd $n$ and there is no energy gap near $E_F$. The even-odd effect on transport does not originate from the difference of the band structures but result from wave function mismatch. In Fig. 3(d) and (f) we present the band structures in the left and right electrodes and a top view of the wave functions of bands π and π*, at the Fermi energy, for 3- and 5-ZαGNRs. Both the π and π* states do not show any specific symmetry. There is no symmetry restriction at the Fermi energy for spin up (down) electrons in the π* (π) state of the left electrode to transport into the π (π*) state of the right electrode in the AP configuration, as indicated by the arrows. On the contrary, as shown in Fig. 3(e) and (g), for 4- and 6-ZαGNR, the π (π*) states of both spins are antisymmetric (symmetric) with respect to the central axis (dash dotted line) due to the transversal geometry symmetry of even-width ZαGNRs. For each spin, the π state in one electrode is orthogonal to the π* states in the other electrode and electrons at $E_F$ cannot tunnel between the two electrodes as indicated by the crossed arrows. This mismatch of the wave functions between π and π* bands results in the conductance gap near $E_F$ in even-width pristine ZαGNRs.

In an even-width edge doped ZαGNR, the impurity bound states can couple the π states in one electrode with the π* states in the other electrode and then can close up the conductance gap at $E_F$ in the AP configuration as shown in Figs. 3(b)-(c). In addition, the conductance becomes spin dependent with a spin-up peak below $E_F$ and a spin-down one above $E_F$. In general, increasing the width of the ZαGNRs makes the spin conductance peaks shift towards or away from $E_F$. Surprisingly, a dip appears in the conductance of odd-width ZαGNRs edge doped by the elements of groups III and V as shown in panels of the first- and third-rows of Fig. 3(b)-(c). Furthermore, the conductance of doped odd-width ZαGNRs appears almost spin degenerate. The common feature in doped ZαGNRs of both even and odd width is that conductance dips appear at energies of the impurity bound states due to the Fano effect, similar to the cases in the P configuration (see Fig. 2). In Fig. 3, we observe a conductance dip near 0.3eV (-0.3eV) in each B (N)-doped ZαGNR.

*D. Tunneling Magnetoresistance*

As discussed above, the conductance may vary greatly between P and AP configurations because the electrons of opposite spins near the Fermi energy are in states of $\pi$ and $\pi^*$ bands, respectively. In transversally symmetric even-width pristine ZαGNRs, the $\pi$ and $\pi^*$ states have different symmetries. Strong even-odd characteristics on the MR should be observed. In table 1, we present the linear conductance and the corresponding MR of pristine and doped n-ZαGNRs for odd $n = 3, 5$ and even $n = 4, 6$ in the AP and P configurations. It is seen that the MR values in pristine 4- and 6-ZαGNRs are in the order of $10^6$ % in contrast to the values of order 10 % or less in pristine 3- and 5-ZαGNRs. This result is similar to that in graphene and silicene nanoribbons [17, 32, 36,37,57]. Interestingly, when ZαGNRs are edge doped by III and V group atoms, the situation reverses: the MR in 4- and 6-ZαGNR decreases by at least five orders of magnitude ($10^5$) while that in 3- and 5-ZαGNRs increases by three-to-four orders of magnitude ($10^3$–$10^4$). Furthermore, the absolute value of the MR increase (decreases) with the width in odd-width (even-width) ZαGNRs. This even-odd effect corresponds to the disappearance of the conductance gap in doped even-width ZαGNRs and the appearance of the conductance dip in odd-width ZαGNRs at the Fermi energy in the AP configuration as shown in Fig. 3. All these phenomena suggest that the transversal symmetry of ZαGNR plays a vital role in determining the electronic and magnetic properties.

Table1. Magnetoresistance (MR) in the pristine and doped 3-, 4-, 5- and 6-ZαGNR.

|         | 3-ZαGNR |         |         | 4-ZαGNR |         |         |
|---------|---------|---------|---------|---------|---------|---------|
| Dopant  | $G^P$ (μS) | $G^{AP}$ (μS) | MR(%) | $G^P$ (μS) | $G^{AP}$ (μS) | MR(%) |
| pristine | 77.3 | 77.1 | 0.2 | 77.2 | $1.8\times10^{-3}$ | $4.4\times10^6$ |
| B-doped | 47.1 | $9.8\times10^{-1}$ | $4.7\times10^3$ | 56.5 | 65.1 | -15 |
| N-doped | 48.9 | 2.5 | $1.9\times10^3$ | 58.5 | 67.1 | -15 |
|         | 5-ZαGNR |         |         | 6-ZαGNR |         |         |
| Dopant  | $G^P$ (μS) | $G^{AP}$ (μS) | MR(%) | $G^P$ (μS) | $G^{AP}$ (μS) | MR(%) |
| pristine | 77.3 | 69.6 | 11 | 77.3 | $2.6\times10^{-3}$ | $2.9\times10^6$ |
| B-doped | 61.9 | $6.0\times10^{-1}$ | $1.0\times10^4$ | 65.9 | 63.5 | 4 |
| N-doped | 63.5 | $8.1\times10^{-1}$ | $7.7\times10^3$ | 67.2 | 65.2 | 3 |

To understand how the doping atom affects the electronic structure and the observed transport properties, we present the spin-dependent projected density of states (PDOS) of the atom at the doping site on the lower edge of the central region. For pristine systems in the P configuration as shown in Fig. 4(a), the spin-up (spin-down) PDOS of the C atom peaks near E=−0.1 eV (0.06 eV), the energy of the edge states (see Fig. 1(d)). When a B (N) atom replaces a C atom, the PDOS peaks of both spins shift to the same energy of the impurity bound states $E$=0.3 (−0.3) eV. The PDOS curves have similar profiles and the spin polarization on the doping site is suppressed. The interaction between the extended edge states and the bound impurity states leads to the Fano conductance dip at $E$=0.3 (−0.3) eV as illustrated in Fig. 2(b)-(c). In addition, the PDOS curves of edge C atoms on the neighbor sites show mixed features of the C atom in pristine systems and the B (N) atom in doped system on the doping site.

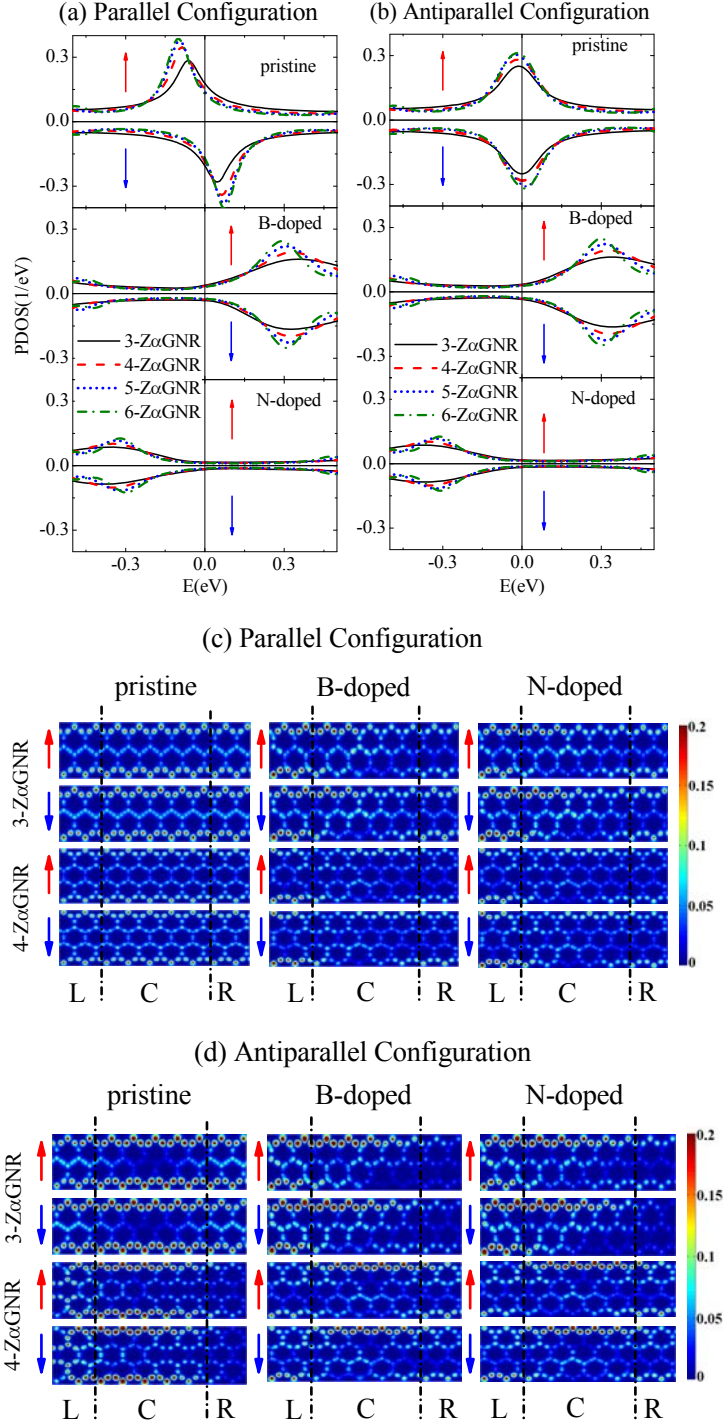

Fig. 4. (Color online) PDOS of the C, B, and N atom at the doping site in pristine, B-doped and N-doped ZαGNRs, respectively, is plotted in the (a) P and (b) AP configurations. Positive (negative) values of PDOS are for the spin up (down). (c) The scattering states at $E_F$ for electrons incident from the left electrode in pristine, B-doped, and N-doped 3- and 4-ZαGNRs in the P configuration. (d) The same as in (c) for the AP

configuration. The color bar represents the amplitude, in unit of Å$^{3/2}$, of the scattering states normalized to the unit current in the electrodes.

In the AP configuration, the left electrode is spin-up polarized and the right one is spin-down polarized. The majority spin in the central region varies gradually from spin-up on the left side to spin-down on the right side. In the studied case as shown in Fig. 1(b), the doping site is closer to the left electrode than to the right one. The C atom on this site in pristine ZαGNRs is slightly spin-up polarized. The spin-down PDOS peak has a higher energy than the spin-up one and the difference between them increases with the width $n$ as depicted in Fig. 4(b). When the C atom is substituted by a B (N) atom the PDOS peaks shift to $E = 0.3$ eV ($E=-0.3$ eV) and have almost the same profile as those in the P configuration.

In Fig. 4(c), we show the scattering states at $E_F$ incident from the left electrode in the P configuration for $n$-ZαGNRs of odd ($n=3$) and even ($n=4$) widths. In pristine systems, the wave functions are extended along the ribbon and confined on the two edges state. When the C atom on the doping site is replaced by a B (N) atom, the wave functions are weakened on the doped edge after scattered by the impurity. Nevertheless, the wavefunctions extend to the right electrode and contribute to the transport. As a result, the linear conductance is only slightly affected in the P configuration as shown in Fig.2.

The scattering states at $E_F$ in the AP configuration are presented in Fig. 4(d). In odd-width pristine 3-ZαGNRs, the wave function extends all the way from the left to the right electrode and is more strongly confined to the edges when passing through the non-magnetized area in the middle of the central region. The conductance of each spin is close to the conductance quantum as illustrated in Fig. 3(a). In contrast, in even-width pristine 4-ZαGNRs, the wavefunction of each spin penetrates through the central region and then vanishes gradually in the right electrode. It is observed that the spin-down wave function decays faster than the spin-up one. When the C atom on the doping site is substituted by a B (N) atom, the wavefunctions behave oppositely for even and odd width number $n$ as shown in the right panels of Fig. 4(d). In 3-ZαGNRs, the impurity blocks the extension of the states and the wave functions decay after passing the impurity which results in the conductance dip at $E_F$ as shown in Fig. 3(b)-(c). In 4-ZαGNRs, however, the impurity breaks the transversal symmetry of the system and the wave function becomes transversally asymmetric. Interestingly, the wave functions of both spins can go around the impurity and extend to the right electrode though they behave in different ways. The linear conductance in this case can be close to the conductance quantum as shown in Fig. 3(b)-(c).

*E．Spin thermoelectric effects*

In Fig. 5(a)-(f) we plot $S_\sigma$ on the left column and the absolute value of $S_C$ and $S_S$ in the right column, as functions of the temperature, for pristine and edge-doped *n*-ZαGNRs (*n*=3, 4, 5, and 6) in the P and AP configurations. We plot the absolute values of $S_C$ and $S_S$ to better compare their magnitude and indicate their zeros, the values at which a pure spin or charge current is produced by the temperature gradient. $S_\sigma$ is linear in temperature at low *T* and become strongly nonlinear as *T* increases[58]. For pristine ZαGNRs in the P configuration, the conductance is equal to $G_0$ near $E_F$. The coefficient $S_\sigma$ is quite small, due to $\tau'_\sigma \approx 0$, and $S_S \approx S_C$ as shown in Fig. 5(a). In doped cases the Seebeck coefficients are greatly enhanced for narrow nanoribbons (*n* = 3 and 4) at high temperature due to the Fano conductance dips at E=±0.3eV. In particular, for the N-doped 5-ZαGNR and 6-ZαGNR, $S_\uparrow$ is negative and $S_\downarrow$ positive in a wider temperature region, as shown in Fig. 5(c). As a result, we note that the maximum of $S_S$ is more than 10 times larger than that of the $S_C$ below room temperature. $S_C$ become zero at T=86 and 301 K for N-doped 6-ZαGNR (see the last panel of Fig. 5(c)), while $S_S$ has a finite value.

In the AP configuration, the transversal symmetry plays an important role in determining the conductance of the ribbons. For pristine odd-width *n*-ZαGNRs (*n*=3 and 5), as shown in Fig. 5(d), the conductance is a constant in a large range of energy and the Seebeck coefficient is negligible up to room temperature. For pristine even-width ZαGNRs, there is a conductance gap near the Fermi energy and the Seebeck coefficients are greatly enhanced. However, the systems are spin degenerate and the spin Seebeck coefficient is near zero.

The doping can bring some interesting results as illustrated in Fig. 5(e) for B-doped and in Fig. 5(f) for N-doped ZαGNRs. At low temperature, the magnitude of the Seebeck coefficients increases linearly as described by Eq.(7) and $S_S$ is always positive. The sign of $S_\sigma$ and $S_C$, however, shows a strong even-odd effect. In odd-width n-ZαGNR ($n = 3, 5$), $S_\uparrow$ and $S_\downarrow$ have the same sign and they are positive in B-doped systems while negative in N-doped systems. In even-width n-ZαGNR ($n=4, 6$), on the contrary, $S_\uparrow$ and $S_\downarrow$ have opposite signs and $S_C$ is negative in B-doped systems while positive in N-doped systems.

At high temperature the Seebeck coefficients become nonlinear versus the temperature. In odd-width n-ZαGNR (n=3, 5), $S_S$ changes sign and a pure charge Seebeck effect ($S_S=0$) occurs below temperature $T=100K$. In addition, usually we have $|S_S|<|S_C|$ except in the temperature region [136, 236] K and [151, 371] K for B-doped 3- and 5-ZαGNR, respectively, as indicated in Fig.5(e). In even-width n-ZαGNR ($n=4, 6$), we have always $|S_S|>|S_C|$ because $S_\uparrow$ and $S_\downarrow$ have opposite signs. This suggests that the edge-doped, even-width ZαGNRs are ideal materials for realizing high-spin-polarization current by using a temperature gradient. More interestingly, we observe that $|S_C|$ can reach to zero at $T=266$ K and 216K in the N-doped 4- and 6-ZαGNR, respectively. This suggest that the N-doped even-width ZαGNRs can be used as the pure spin current generator by a temperature gradient and might be valuable to thermo-spintronics.

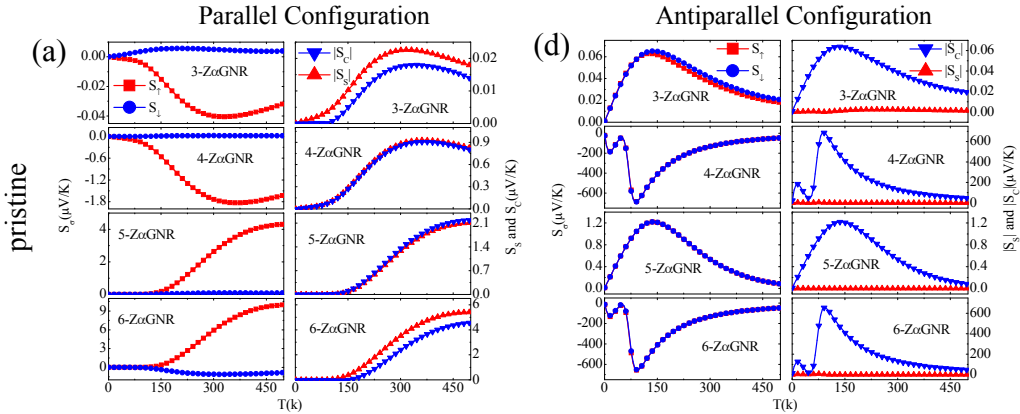

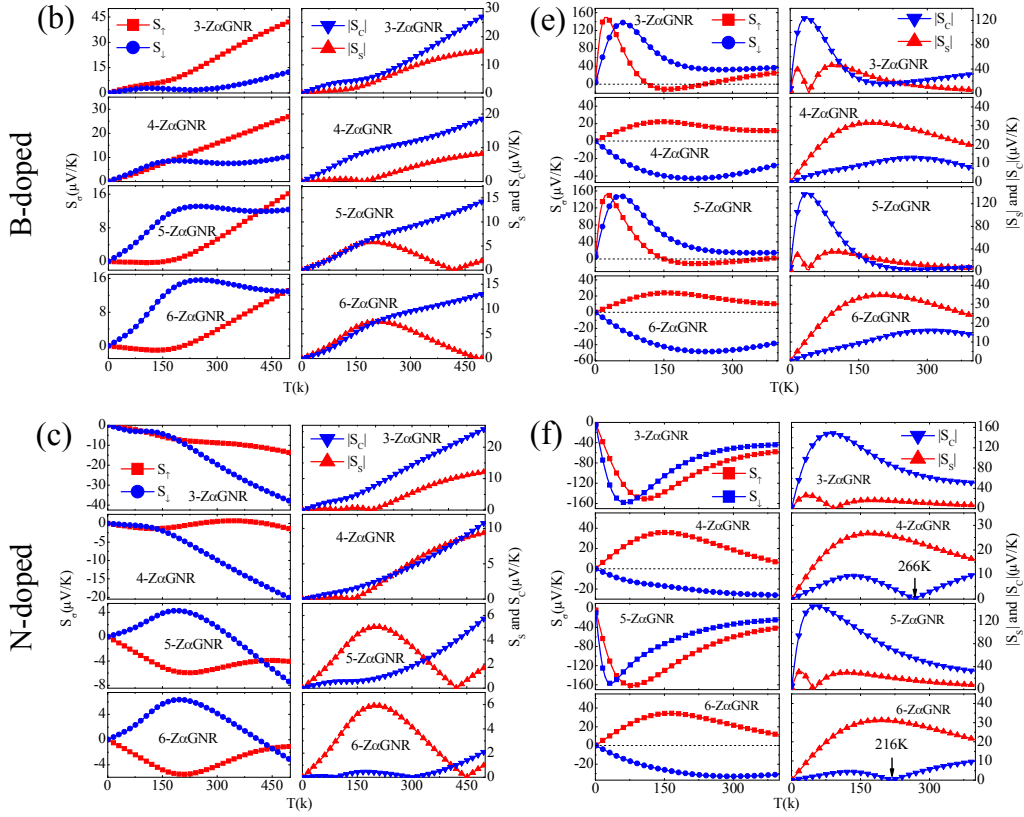

Fig. 5 (Color online) Spin-dependent ($S_\sigma$) Seebeck coefficients (left column) and the absolute value of charge ($|S_C|$), and spin ($|S_S|$) Seebeck coefficients (right column), as functions of temperature $T$ in (a) pristine, (b) B-doped, and (c) N-doped n-ZαGNRs. The left and right panels are for the P and AP configuration, respectively.

The even-odd effect on the Seebeck coefficients observed in doped ZαGNRs has similar origin as the effect on the magnetoresistance. The strong spin Seebeck effect in doped even-width ZαGNRs with AP configuration can be understood from the conductance spectra shown in Fig. 3(b)-(c). There is a spin-up (spin-down) conductance peak just below (above) the Fermi energy which results in sharp but opposite slopes of the conductance spectra at the Fermi energy for opposite spins. From Eq. (7) we then have positive $S_\uparrow$ and negative $S_\downarrow$ and $|S_S|>|S_C|$.

## 4. Summary

We studied transport properties of pristine and edge-doped ferromagnetic ZαGNRs in the parallel and antiparallel electrode configurations, using the density functional theory combined with the non-equilibrium Green's function method. We found that pristine ZαGNRs exhibit strong width-dependent transport characteristics in the antiparallel configuration. This feature originates from the different wave function symmetries of electrons near the Fermi level between odd- and even-width ZαGNRs. In edge-doped, even-width ZαGNRs the dopant breaks the symmetry of the system and the linear conductance recovers to almost $G_0 = e^2/h$ for both spins. On the contrary, in edge-doped, odd-width ZαGNRs the dopant blocks electronic transport via the edge states, the conductance decreases to almost zero. The swap of linear conductances between odd- and even-width ZαGNRs after being edge doped leads to the drastic change of the corresponding magnetoresistance by several orders of magnitude. This results can be very useful in manipulating magnetoresistance in devices.

In addition, due to the same mechanism, we found that the thermoelectric properties also show a strong odd-even effect in the antiparallel configuration. For doped even-width ZαGNRs, the magnitude of the charge Seebeck coefficient is much smaller than that of the spin Seebeck one below room temperature and can even become zero at specific temperatures. This suggests that edge-doped ZαGNRs can be used to make thermospin devices which create pure spin current upon applying temperature gradients.

**Acknowledgments**

We thank Lei Liu and Eric Zhu for assistance in numerical calculations using the NanoDcal software. This work was supported by the National Natural Science Foundation in China (Grant Nos. 11074182, 91121021, and 11247028) and by the Canadian NSERC Grant No. OGP0121756.